%% file: template.tex
\documentclass[journal]{vgtc}
\ifpdf%                                % if we use pdflatex
  \pdfoutput=1\relax                   % create PDFs from pdfLaTeX
  \usepackage{graphicx}                % allow us to embed graphics files
  \DeclareGraphicsExtensions{.pdf,.png,.jpg,.jpeg} % for pdflatex we expect .pdf, .png, or .jpg files
\else%                                 % else we use pure latex
  \usepackage{graphicx}                % allow us to embed graphics files
  \DeclareGraphicsExtensions{.eps}     % for pure latex we expect eps files
\fi%

\graphicspath{{figures/PDF/}{images/}{./}} % where to search for the images

\usepackage{microtype}                 % use micro-typography (slightly more compact, better to read)
\PassOptionsToPackage{warn}{textcomp}  % to address font issues with \textrightarrow
\usepackage{textcomp}                  % use better special symbols
\usepackage{mathptmx}                  % use matching math font
\usepackage{times}                     % we use Times as the main font
\usepackage{cite}                      % needed to automatically sort the references
\usepackage{tabu}                      % only used for the table example
\usepackage{booktabs}                 
\usepackage{paralist}
\usepackage{enumitem}
\setlist{topsep=0pt, leftmargin=*}
\usepackage{graphicx} %插入图片的宏包
\usepackage{float} %设置图片浮动位置的宏包
\usepackage{subfigure} %插入多图时用子图显示的宏包
\usepackage{color}
\usepackage{wrapfig}
\usepackage{xcolor}
\usepackage{booktabs}
\usepackage{setspace}
\usepackage{stfloats}
\usepackage{amsmath}
\usepackage{soul}
\usepackage{amssymb}

\definecolor{brandeisblue}{rgb}{0.0, 0.44, 1.0}
\definecolor{addgreen}{HTML}{138a07}
\definecolor{refer_color}{RGB}{200,100,100}

\newcommand{\add}[1]{{\color{black} #1}}
\newcommand{\mgv}{{MGV}}
\newcommand{\quo}[1]{\textit{``#1''}}

\newcommand{\subtitle}[1]{{\noindent\textbf{#1}}}

\setenumerate[1]{itemsep=1pt,partopsep=1pt,parsep=\parskip,topsep=1pt}
\setitemize[1]{itemsep=1pt,partopsep=1pt,parsep=\parskip,topsep=1pt}
\setdescription{itemsep=1pt,partopsep=1pt,parsep=\parskip,topsep=1pt}

\preprinttext{To appear in IEEE Transactions on Visualization and Computer Graphics.}

\ieeedoi{xx.xxxx/TVCG.201x.xxxxxxx}

\onlineid{1335}

 \vgtccategory{Research}
 \vgtcpapertype{algorithm/technique}

\title{MetaGlyph: Automatic Generation of \\ Metaphoric Glyph-based Visualization}

\author{Lu Ying, Xinhuan Shu, Dazhen Deng, Yuchen Yang, Tan Tang, Lingyun Yu, Yingcai Wu}

\authorfooter{
\item L. Ying, D. Deng, Y. Yang, Y. Wu are with the State Key Lab of CAD\&CG, Zhejiang University, Hangzhou, China. Y. Wu is also with the Alibaba-Zhejiang University Joint Research Institute of Frontier Technologies. 
E-mail: \{yingluu, dengdazhen, yyc\_yang, ycwu\}@zju.edu.cn. 
\item X. Shu is with Department of Computer Science and Engineering, The Hong Kong University of Science and Technology, Hong Kong, China. E-mail: xinhuan.shu@connect.ust.hk.
\item T. Tang is with School of Art and Archaeology, Zhejiang University, Hangzhou, China. E-mail: tangtan@zju.edu.cn.
\item L. Yu is with Department of Computing, Xi’an Jiaotong-Liverpool University, Suzhou, China. Email: Lingyun.Yu@xjtlu.edu.cn.
\item Yingcai Wu is the corresponding author.
}

\shortauthortitle{Biv \MakeLowercase{\textit{et al.}}: Global Illumination for Fun and Profit}
\abstract{
  \input{paper/abstract.tex}
} % end of abstract


\CCScatlist{ % not used in journal version
 \CCScat{K.6.1}{Management of Computing and Information Systems}%
{Project and People Management}{Life Cycle};
 \CCScat{K.7.m}{The Computing Profession}{Miscellaneous}{Ethics}
}

\teaser{
  \centering
  \includegraphics[width=\linewidth]{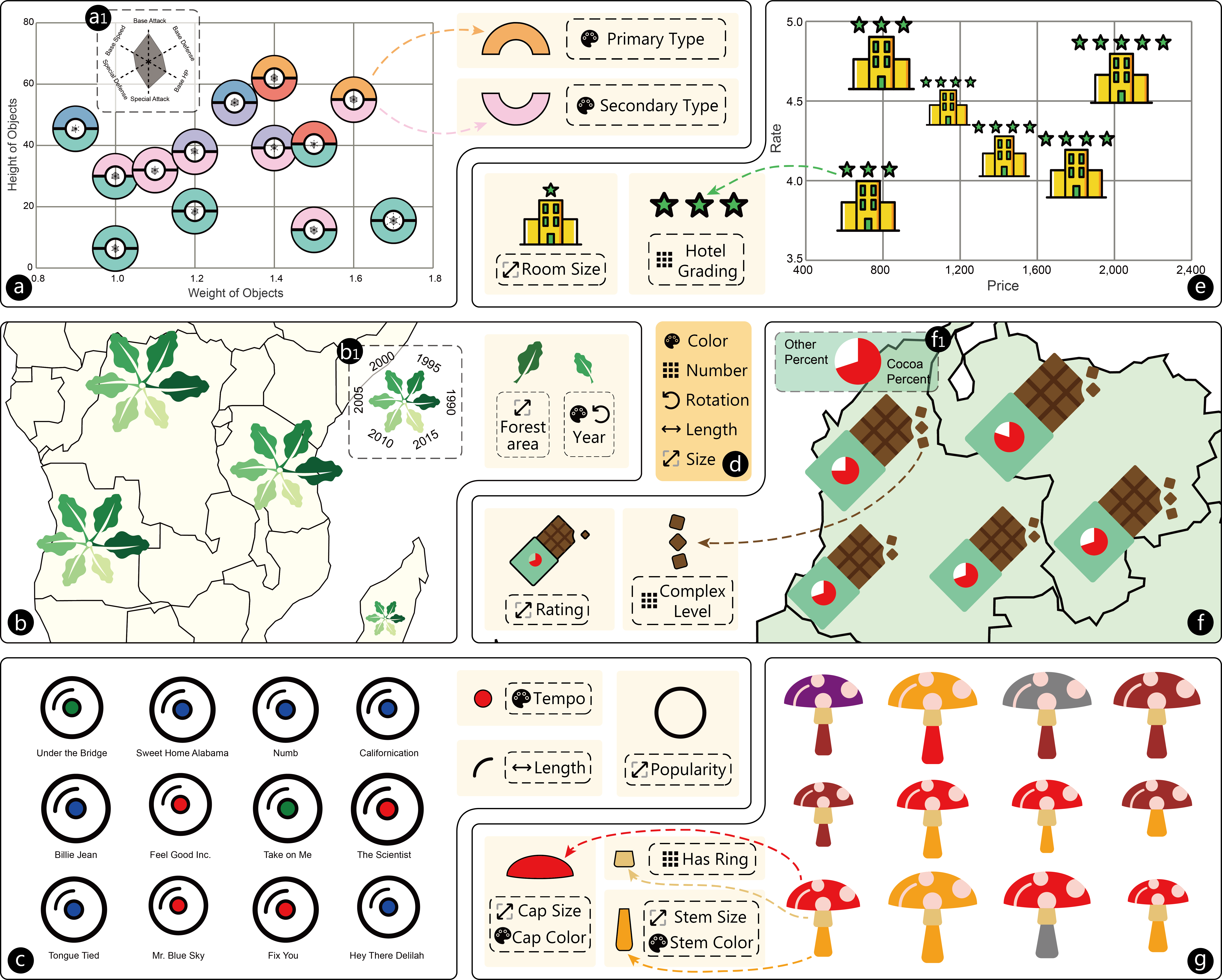}
  \caption{
  Example MGVs generated by MetaGlyph:
  (a) several attributes of different pokemons;
  (b) forest area changes in different countries from 1995 to 2020;
  (c) display of various CDs;
  (e) multiple hotels' information arranged by price (x) and rate (y);
  (f) chocolates in different ratings placed on a map;
  (g) disparate mushrooms.
  The legends in (a1)(b1)(f1) are also generated by MetaGlyph.
  The center boxes with yellow backgrounds illustrate the data mappings for each glyph.
  Encoding channels are represented in icons and (d) shows the annotation.
  Dashed arrows are used to associate visual elements with the resulting visualizations.
  }
  \label{fig:teaser}
}

\vgtcinsertpkg

\begin{document}
\begin{spacing}{0.97}
\maketitle
\setuldepth{Score}

\input{paper/1_intro.tex}
\input{paper/2_related_work.tex}
\input{paper/3_Design_of_MetaGlyph}

\end{spacing}
\begin{spacing}{0.965}
\input{paper/4_Generation_Engine}
\input{paper/5_MetaGlyph}

\input{paper/6_Evaluation}
\input{paper/7_Discussion}
\end{spacing}

% \begin{figure*}[!t]
%   \centering
%   \includegraphics{gallery}
%   \caption{
%     Examples created by GlyphCreator.
%     }
%   \label{Fig.gallery}
% \end{figure*}

% \input{paper/discussion.tex}

\acknowledgments{
The work was supported by NSFC (62072400) and the Collaborative Innovation Center of Artificial Intelligence by MOE and Zhejiang Provincial Government (ZJU).
}

\bibliographystyle{abbrv-doi}

\newpage
% \normalem
\bibliography{template}
\end{document}

%% file: paper/abstract.tex
Glyph-based visualization achieves an impressive graphic design when associated with comprehensive visual metaphors, which help audiences effectively grasp the conveyed information through revealing data semantics.
However, creating such metaphoric glyph-based visualization (MGV) is not an easy task, as it requires not only a deep understanding of data but also professional design skills. 
This paper proposes MetaGlyph, an automatic system for generating \mgv{}s from a spreadsheet.
To develop MetaGlyph, we first conduct a qualitative analysis to understand the design of current \mgv{}s from the perspectives of metaphor embodiment and glyph design.
Based on the results, we introduce a novel framework for generating \mgv{}s by metaphoric image selection and an \mgv{} construction.
Specifically, MetaGlyph automatically selects metaphors with corresponding images from online resources based on the input data semantics. 
We then integrate a Monte Carlo tree search algorithm that explores the design of an \mgv{} by associating visual elements with data dimensions given the data importance, semantic relevance, and glyph non-overlap. 
The system also provides editing feedback that allows users to customize the \mgv{}s according to their design preferences.
We demonstrate the use of MetaGlyph through a set of examples, one usage scenario, and validate its effectiveness through a series of expert interviews.

%% file: paper/1_intro.tex
% !TEX program = pdflatex
\section{Introduction}
Glyph-based visualization serves as an effective method for encoding multi-dimensional data~\cite{MANSOOR2021ARGUS}.
However, glyph designs can also be complex due to the increasing number of data dimensions, leading to comprehension problems.
% understanding the glyph becomes more difficult. 
Accordingly, visual metaphors are actively used to draw data-driven glyphs with representative and familiar appearances related to the data~\cite{Li2015MetaphoricTransfer}.
% to align them with familiar schema
We have seen wide adoption of metaphoric glyphs in various domains, such as sports~\cite{legg2012matchpad, polk2014tennivis}, urban application~\cite{liu2017smartadp, deng2022compass}, and blockchain~\cite{zhong2020silkviser}.
Studies have also shown that appropriate metaphors can help people understand glyphs quickly and accurately~\cite{Cunha2018ManyFacedPlot, Fuchs2017GlyphReview, Li2015MetaphoricTransfer}.
% As a result, we've seen wide adoption of metaphoric glyphs in various domains, such as sports~\cite{legg2012matchpad, polk2014tennivis}, urban~\cite{liu2017smartadp, deng2022compass}, and news media~\cite{cao2016targetvue}.
% 创建困难
However, incorporating metaphors in glyph designs is not an easy task. 
Designers have to balance various factors, such as the expressiveness of the visual representations and the effectiveness of data mappings.

Many advanced visualization authoring tools have been developed to facilitate the creation of glyph-based visualizations~\cite{Kim2017DDG, Xia2018DataInk, Ren2019Charticulator, ZhangSBC20DataQuilt, Ying2022GlyphCreator, Brehmer2022Diatoms}.
However, it is difficult to balance automation and customization in the creation process.
\add{For example, GlyphCreator~\cite{Ying2022GlyphCreator} and Diatoms~\cite{Brehmer2022Diatoms} consider basic geometry or limited shapes and do not have explicit support to create metaphoric glyphs with semantic relevance.
Other manual authoring tools allow users to craft the graphical elements from scratch through sketching~\cite{Xia2018DataInk} or interactions~\cite{Kim2017DDG}, which are powerful in customization.
% Although such tools provide convenience for users, they cannot balance between visualization diversity and ease-of-use.
% Some tools only support basic geometry~\cite{Ying2022GlyphCreator} or limited shapes~\cite{Brehmer2022Diatoms} and do not support the creation of glyphs with metaphors.
% Some tools support a variety of elements as metaphors but require the users to draw from scratch, either by sketching~\cite{Xia2018DataInk} or using interactions~\cite{Kim2017DDG}.
Such creation is laborious, and the quality of the final glyphs is highly dependent on the user's design experience and expertise.
% 现在网上有素材库 -》降低定制化成本[]
Using online resources may lower the cost of customization and simplify the creative process~\cite{ZhangSBC20DataQuilt}.
% Recently, some tools were proposed to use online images for glyph creation, such as DataQuilt~\cite{ZhangSBC20DataQuilt}.
% Users should upload images on their own and further create glyphs based on them.
% However, users may easily get lost in the vast amount of image sources and design choices.
However, users need to manually select and upload the image source online without an image library.}

% 自动借助资源
For better results, we aim to automatically generate metaphoric glyph-based visualization (MGV) using online sources.
% 为什么要自动
We attempt to ease the creation of \mgv{} for general users who need to encode multi-variant data.
However, two obstacles emerge from the process:
% \vspace{2pt}
\begin{itemize}
    \item \textit{It is unclear how metaphors can be embedded into the glyph-based visualization design.}
    In practice, a great number of visualizations~\cite{zhong2020silkviser, xie2021passvizor, xu2019clouddet, cui2021VineMap} have adopted metaphors to represent data.
    Existing studies~\cite{Borgo2013GlyphbasedVisualization, wenskovitch2018systematic} have recognized that metaphor is a good design strategy to facilitate glyph understanding.
    However, a systematic review of these designs to guide the creation of \mgv{} is lacking.
    Detailed and practical designs for \mgv{} have not been proposed yet.
    % Researchers~\cite{Borgo2013GlyphbasedVisualization, wenskovitch2018systematic} have proposed general design guidelines, including metaphoric designs that facilitate data understanding.
    % However, they do not elaborate on the detailed design ideas of \mgv{}.
    % A great number of visual analytic systems~\cite{zhong2020silkviser, xie2021passvizor, xu2019clouddet} have adopted metaphors in their design.
    % Focused on the data in a specific field, they introduce a design to solve their question.
    % However, these experience cannot serve as helpful guides for a general \mgv{} design.
    
    % \item \textit{How to transform the human design process into a framework that a machine can imitate and automate is difficult.}
    \item \textit{It is difficult to generate an \mgv{} without the involvement of human intelligence.}
    % 具体流程未知
    This generation process involves a series of design decisions, for instance, selecting an appropriate metaphor design and binding data with various elements in the metaphor.
    It requires comprehensive considerations of the whole process to streamline the production. 
    % It requires a technique that can consider critical decisions comprehensively and generate \mgv{}s as a result.
    % 之前没有这样子的工作，需要提及吗
    Although increasing work has been conducted on glyph-based visualization, the automatic method for designing and generating an \mgv{} has received less attention from the community.
\end{itemize}
% \vspace{2pt}
To address these challenges, we propose an \mgv{} generation framework according to a qualitative analysis established on a collection of \mgv{} examples.
We systematically reviewed 50 examples from publications and websites to explore the \mgv{}s' designs.
% in different stages.
The analysis results provide guidance for understanding a metaphor within a glyph-based visualization for the first challenge.
Informed by the results, we design and implement MetaGlyph, a proof-of-concept system that allows users to generate \mgv{}s automatically by importing a spreadsheet.
We find appropriate metaphoric images online for the data and assess how well the images match the input data in the following steps:
First, the images are decomposed into a list of visual elements.
Given the visual elements and different data dimensions, 
%we try to find a satisfactory mapping.
we then formulate the mapping problem as a tree search question and introduce a Monte Carlo tree search (MCTS) algorithm to explore the mapping space.
Finally, we utilize criteria to estimate the quality of \mgv{}s and select the best \mgv{} based on the rewards.
% A reward function is utilized to estimate the quality.
% We select \mgv{}s with high scores based on the value as the final result.
MetaGlyph also incorporates an interface for users to refine the \mgv.
The main contribution can be summarized as follows:
\begin{itemize}[noitemsep]
    \item We conduct a qualitative analysis to understand the design of state-of-the-art \mgv{}s from various stages.
    \item We propose a novel framework by selecting metaphoric images and constructing \mgv{}s.
    We also introduce a method to estimate the quality of an \mgv{} in the framework.
    \item We develop MetaGlyph, a mixed-initiative system for creating \mgv{}s automatically. 
    We demonstrate the usage of MetaGlyph through a usage scenario and validate its usability through expert interviews.
\end{itemize}

% 选隐喻+图片
% First, among all options, we need to choose a representational and familiar metaphor for all the input data.
% For each particular metaphor, there are a large collection of images on the web.
% How to find an appropriate image for data is unknown.
% Second, encoding data into a suitable image is not an easy task.
% We need to decompose a picture into different elements and bind data with elements of different irregular shapes.
% Moreover, we also need to consider that the transformed form of each element and each glyph cannot conflict with the overall data presentation.
% For example, when leveraging the car metaphor, the ultimate visualization cannot cause too much overlap between glyphs.

%% file: paper/2_related_work.tex
% !TEX program = pdflatex
\section{Related Work}

We summarize prior studies that have covered metaphor-based designs, glyph-based visualizations, and currently available authoring tools.

\subsection{Metaphor-based designs}
% 语言学和可视化中的概念
In linguistics, metaphor, analogy, and simile are three basic elements in language~\cite{Norvig1985Metaphors}.
A metaphor is a word or a phrase that compares one thing to another to make a description more intuitive, like ``\textit{All the world is a stage}''~\cite{gilbert1989evaluation}.
% analogy simile 关系
Similes create a comparison using ``like'' or ``as''~\cite{gilbert1989evaluation}. 
A well-known example of a simile is ``\textit{Life is like a box of chocolates.}''
The analogy is a comparison between things that have similar features~\cite{gilbert1989evaluation}.
An example of an analogy is ``\textit{Black is to white as on is to off}''.
\add{In visualization, researchers do not explicitly differentiate these concepts, and the word ``metaphor'' is used to depict the case of interpreting complex information via familiar and concrete objects~\cite{Li2015MetaphoricTransfer}.}
By allowing users to maximize their experience and knowledge, metaphors make it easier for users to understand the underlying data.

% 实验证明
Previous studies have suggested that metaphors promote data comprehension~\cite{risch2008role, Fuchs2017GlyphReview}.
A well-known example is Chernoff faces~\cite{chernoff1973use}, which maps one data value to one face character like the eyebrows' angle or the nose's size.
Later in two quantitative experiments, Flury et al.~\cite{flury1981graphical} and Jacob~\cite{jacob1978facial} found that face glyphs outperform other visual designs like polygons and digits.
Researchers have proved that, compared with glyphs unrelated to data, some metaphor-based glyphs outperform others in accuracy and efficiency through quantitative experiments, such as car glyphs~\cite{surtola2005effect} and clock glyphs~\cite{fuchs2013evaluation}.
Chau et al.~\cite{chau2011visualizing} found that a combined design that displays glyphs and numbers together performs better in adopting a flower metaphor.
Fuchs et al.~\cite{fuchs2016leaf} recently introduced a leaf glyph based on a natural metaphor and proved its effectiveness in illustrative storytelling.
Dasu et al.~\cite{dasu2018organica} proposed an organic metaphor to interpret conditional co-occurrences and verified the effectiveness of complex tasks.

% 所以用的多

\add{
The concept of metaphor in visualization has a long history.
In the 1920s, Otto Neurath and Gert Arntz~\cite{neurath1973vienna} invented the
`Vienna Method of Pictorial Statistics', which was renamed `ISOTYPE (International System Of TYpographic Picture Education)' in the late 1930s.
They designed a lot of pictographs using semantically relevant icons.
% 语义相关

% 感觉这个话不能这么写
Metaphors have been widely used to visualize different data.}
% Many visual analytic systems have employed metaphors in different layouts.
Metaphors like clock~\cite{el-assady2019visual}, wheel~\cite{alsallakh2012reinventing}, and radar~\cite{wu2018visual} are adopted to present radial layouts. 
Spatial metaphors express the relation ``$proximity \approx similarity$''~\cite{montello2003testing}.
Ropinski et al.~\cite{ropinski2011survey} recommended 3D metaphoric glyphs to visualize spatial multivariate medical data for the attentive phase.
Using Fermat's spirals, Lei and Zhang~\cite{lei2011visual} designed the galaxy visualization to display financial time serials.
Matchpad~\cite{legg2012matchpad} used metaphoric pictograms, which are easy to learn, remember and guess.
\add{Setlur and Mackinlay~\cite{setlur2014automatic} generated scatterplots with semantically-relevant icons to replace traditional data points automatically.
Users are kept informed by the semantic information during analysis.}
TenniVis~\cite{polk2014tennivis} proposed a novel glyph for individual point outcomes in a tennis match inspired by the needle gauge.
SmartAdp~\cite{liu2017smartadp} designed a novel dashboard-like glyph to represent a solution for billboard placements.
\add{Coelho and Mueller~\cite{coelho2020infomages} created Infomages, which utilized thematic images to develop a data chart. 
A relevant image help users interpret the data.}
Recently, Compass~\cite{deng2022compass} introduced a compass glyph to facilitate the in-depth understanding of urban problems.

% well accepted
% tool -> promote -> design
% 目的

Although these works have demonstrated the great advantages of metaphors in the visualization from different perspectives, none of them have proposed an automatic way to generate metaphoric glyphs.

\subsection{Glyph-based visualization and authoring}

% Glyph-based visualization has become prevalent in presenting multivariate data~\cite{Borgo2013GlyphbasedVisualization}
\add{Glyph-based visualization has become prevalent in visualization journals~\cite{Borgo2013GlyphbasedVisualization, ZikunJoV, wang2022AFExplorer} and celebrated collections (e.g., Dear Data~\cite{lupi2016dear}).
It performs well especially for multivariate data~\cite{Borgo2013GlyphbasedVisualization}}.
However, it is not easy to design and generate a glyph-based visualization.

% 强调设计上的理论研究
Thus, many theoretical researchers have focused on how to design glyphs.
Ward~\cite{Ward2008Multivariate} discussed the process and issues of glyph generation, including mapping data to graphics attributes and layouts.
Borgo et al.~\cite{Borgo2013GlyphbasedVisualization} drew the link between basic concepts in semiotics and glyph-based visualization and summarized existing design guidelines and techniques.
Recently, Fuchs et al.~\cite{Fuchs2017GlyphReview} provided an overview of glyph types and design characteristics by reviewing experimental studies.
% Cunda et al.~\cite{cunha2018manyfaced} focused on emoji systems and presented an approach for generating glyphs automatically.

Furthermore, researchers have proposed many authoring tools to ease the difficulties of creating glyphs.
Ribarsky et al.~\cite{Ribarsky1994Glyphmaker} introduced Glyphmaker, which allows non-expert users to customize data glyphs.
Kim et al.~\cite{Kim2017DDG} proposed Data-Driven Guides for Information Graphics, a system that can also create glyphs via interaction.
Xia et al.~\cite{Xia2018DataInk} developed DataInk, which aims at author glyphs through freedom sketching.
Ren et al.~\cite{Ren2019Charticulator} presented Charticular, an authoring tool focused on layouts between glyphs.
Chen et al.~\cite{chen2020marvist} took the first step in creating glyph-based visualization in Augmented Reality environments using mobile devices.
Besides creating from scratch, DataQuilt~\cite{ZhangSBC20DataQuilt} adopted real images for both inspiration and a resource of visual elements for data binding.  
\add{On the other hand, while scholars have developed several tools to help users create glyphs, users still need a lot of manual operations in the system.
The glyph quality highly depends on the user's design experience and expertise.}
Therefore, scholars have proposed some automatic systems to simplify the process recently.
Ying et al.~\cite{Ying2022GlyphCreator} aimed at circular glyphs and introduced GlyphCreator based on an example-based method.
Brehmer et al.~\cite{Brehmer2022Diatoms} developed Diatoms, a technique for inspiring glyph design through a sample-based generative process.
\add{However, automatic systems only support basic geometry or limited shapes.
Unlike regular shapes, metaphors serve as effective methods to help users understand data.}
Thus, we opt to generate glyphs with metaphors that are not supported by existing systems.

% 总结

%% file: paper/3_Design_of_MetaGlyph.tex
\section{Design of the MetaGlyph System}

In this section, we introduce the design of the MetaGlyph system.
To gain a better understanding of MGV design, we surveyed previous work and conducted a qualitative analysis. 
% To help us understand how human designers create metaphoric glyph-based visualization, we first survey on a corpus of \mgv{} and then give the formal definition.
Based on the findings, we proposed the design considerations (DCs) for the MetaGlyph system.

\subsection{Data Collection}\label{section: prestudy}

% 收集
To better understand how metaphors are embodied in glyph-based visualization, we examined the practice from both academic literature and online design communities. 

% We further expand the corpus from online design community.
% To better understand how people create metaphoric glyph-based visualization, we look for answers from examples.
% Considering the professionalism of the design, we gave priority to examples in the literature.
We collected examples from top visualization conferences and journals (IEEE VIS and TVCG) using the keywords \textit{``metaphor''} and \textit{``glyph''} and found 221 papers.
\add{We manually examined all papers to ensure the existence of MGVs in the paper. Specifically, we checked whether both keywords \textit{``metaphor''} and \textit{``glyph''} appeared together to describe a metaphoric glyph.
% We need to ensure the existence of MGVs in the paper.
Some papers were excluded. 
For instance, they might mention \textit{``metaphor''} and \textit{``glyph''} in the related work for two different works, respectively. 
% in some papers, \textit{``metaphor''} is mentioned in the related work  while \textit{``glyph''} is mentioned in another work.
}
% For the valid paper, we analyzed what metaphor is used and how the metaphor is embedded in the glyph design. 
% We found that flowers~\cite{Albo2016radar, NanCao2012whisper} and clocks~\cite{WeiweiCui2011textflow, Wu2020visual} are popular metaphors among researchers since the metaphor corresponds to the data structure.
% Some papers introduced novel metaphors like coin~\cite{Zhong2020silkviser} or ping-pong table~\cite{Wu2018ittvis}, which is related to the topic of data.
As a result, we collected 20 examples in the literature as the initial corpus.

To further expand the diversity, we collected more examples from creative websites (e.g.,\textit{Behance} and \textit{Pinterest}) with keywords such as \textit{``information visualization''}  \textit{``metaphor''} and \textit{``glyph''}.
We adopted an initial filtering standard of MGV based on our current corpus.
Then, we collected the chosen examples as a new part of our corpus.
Finally, a total of 50 \mgv{}s were collected as our corpus.

\begin{figure}[htbp] 
  \centering %图片居中
  \includegraphics{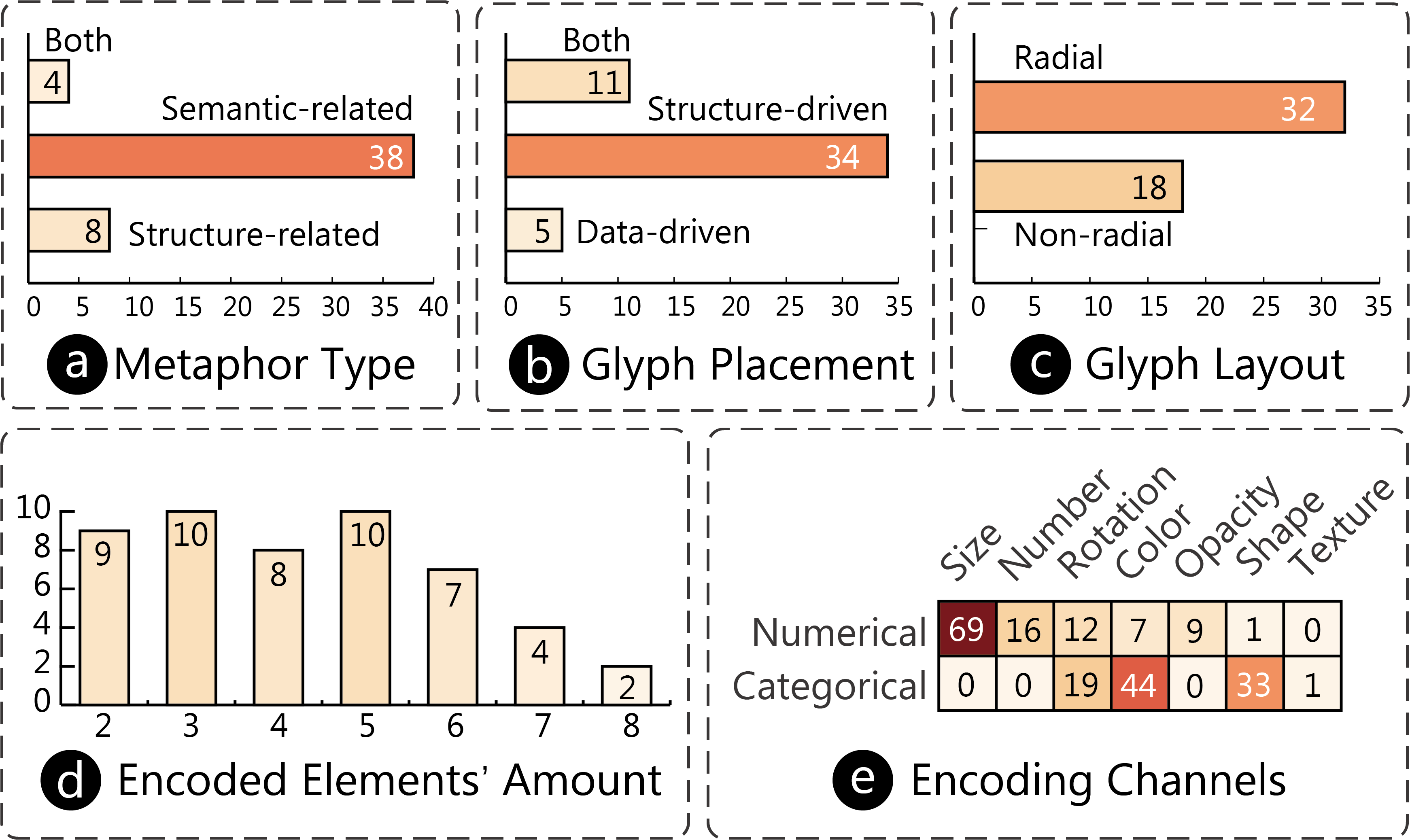}
  \caption{
    The number of \mgv{}s (a) in different metaphor types, (b) in different glyph placements, and (c) in different glyph layouts. 
    % based on qualitative analysis.
    The frequency of (d) amounts of different encoded elements and
    (e) different encoding channels for numerical and categorical data types in our corpus.}
  \label{Fig:analysis} %用于文内引用的标签
\end{figure}

\subsection{Qualitative Analysis}\label{section: analysis}

To further understand the design of an \mgv, we conducted a comprehensive analysis for all examples in our corpus.
In general, we analyzed all \mgv{}s in three stages by decomposing them step by step.
% \vspace{2pt}
\begin{description}[itemindent=-1.5em]
    \item[S1]
    We find out how metaphors are embodied in an \mgv{} and analyze the pattern of the glyph placement based on the overall design.
    \item[S2]
    We drill down into a single glyph and consider its layout.
    \item[S3]
    We aim at different visual elements that compose one glyph.
\end{description}
% \vspace{2pt}
Three authors went through our corpus and independently analyzed all examples following the above stages.
All disagreements were resolved through iterative discussions.

\vspace{1pt}
\subtitle{S1:} At this stage, we focus on \mgv{}s' general design to understand how designers embody the metaphor and place all glyphs in the \mgv{}s.

\begin{itemize}
    \item \textbf{Metaphor Type.}
    In our corpus, we find that designers use metaphors with respect to data properties and generally divide all examples into two conditions: semantic-related and structure-related.
    The former indicates the metaphor is related to the topic of data, for example, a coin representing transaction data~\cite{zhong2020silkviser}, and a dashboard representing speed data~\cite{liu2017smartadp}.
    In this condition, designers commonly use a metaphor within one glyph design.
    The latter illustrates that the metaphor is related to the data structure.
    For example, researchers used blooming flowers in a glyph-based visualization to express hierarchical data~\cite{el-assady2019visual}.
    For data about model optimization, researchers may choose the clock metaphor~\cite{zhao2014fluxflow} since it is time-related.
    In this condition, the metaphors are mainly embodied in the visualization layout.
    Moreover, some designs are both semantic-related and structure-related.
    The concrete amounts of different metaphor types in our corpus are shown in \autoref{Fig:analysis}(a).
    % In this paper, we define an \mgv{} covering both two conditions.
    Finally, the definition of a valid \mgv{} is:
    \textbf{using a visual design that suggests a particular association or similarity with data}~\cite{kogan1980understanding}. 

    \item \textbf{Glyph Placement.}
    In our corpus, we code the glyph placement of \mgv{}s into two groups according to Ward's theory~\cite{Ward2008Multivariate}:
    data-driven and structure-driven.
    Data-driven placements correspond to glyphs that are placed based on data values.
    Some data values are directly used as the x- or y-value. 
    Others need computations (e.g., projection space) to derive the position.
    Glyphs are usually placed in a Cartesian coordinate system in a data-driven group.
    The structure-driven group assumes that the data have structural characteristics.
    It is also a method to present metaphors.
    Some placements are based on a specific object, such as a tree for hierarchical data~\cite{Jang2016MotionFlow} and a map for geospatial data~\cite{proma2021cleanairnowkc}, or a timeline for time series data~\cite{rind2013timebench}.
    Designers adopt metaphors in structure-driven placements, such as a map with landscape and a clock for data ordered by temporal data.
    Other metaphors use a typical ordering relationship based on categorical information.
    Glyphs may be arranged evenly between left and right or located with an organization considering non-overlapping.
    In some cases, designers adopt two placements together for better visualization.
    \autoref{Fig:analysis}(b) indicates the statistics on the corpus.
\end{itemize}
\vspace{1pt}
\subtitle{S2:} We then drill down into the design of one glyph.
We focus on different glyph layouts in this stage.

\begin{itemize}[noitemsep]
    \item \textbf{Glyph Layout.}
    As glyphs are frequently designed in a radial structure~\cite{Ying2022GlyphCreator}, we discuss the glyph layout in two groups: radial and non-radial.
    \autoref{Fig:analysis}(c) displays the used frequency of two groups in our corpus.
    \add{A radial glyph is a glyph whose elements are organized on a polar coordinate system.
    Each element shares the same origin.
    Most elements have a radial shape, like a circle and a sector.
    Non-radial glyphs can be placed in a Cartesian coordinate system.}
    Notably, some glyphs present a linear structure, that is, elements in such glyphs are arranged vertically or horizontally.
    Others are arranged relatively freely, such as to compose a specific object like a car~\cite{surtola2005effect}. 
    % Others have an arbitrary layout, whose elements are in a random arrangement, such as a car glyph~\cite{surtola2005effect}. 
  \end{itemize}

\vspace{1pt}
\subtitle{S3:} 
A glyph is composed of different visual elements encoded by different data dimensions.
In the last stage, we focus on the visual elements and discuss some findings of data mapping.

\begin{itemize}
    \item \textbf{Visual Element.}
    Ying et al.~\cite{Ying2022GlyphCreator} divided all visual elements in a circular glyph into four categories: chart, shape, icon, and text.
    % 需要讲清楚为什么不要icon和text吗
    Given the peculiarity of metaphor, we mainly consider two of these categories: shape-level and chart-level.
    The shape-level element refers to different shapes, including basic geometry (e.g., circles, polygons) and complex shapes (e.g., leaves).
    % Such elements usually share the same color.
    A chart-level element is a variant chart within a glyph, which is also a unit of shape-level elements.
    We integrate these shape-level elements because the unit (e.g., a pie chart) conveys more information than a single component (e.g., several sectors).
    In our corpus, designers adopt pie charts, donut charts, star plots, heatmap, and boxplots when designing glyphs.
    \item \textbf{Element Number.}
    % 只能传递有限的信息
    The information conveyed by one glyph is limited.
    A glyph can better represent data in 2 to 4 dimensions~\cite{Borgo2013GlyphbasedVisualization}.
    According to the statistics from our corpus, the frequent amount of encoded elements of a metaphoric glyph is between 2 and 6 (\autoref{Fig:analysis}(d)).
    \item \textbf{Data Mapping.}
    % 因此mapping很重要
    The mapping relationship of the data and elements is important for presenting the final visualization.
    For a given data dimension and a given element, the encoding channel is mainly determined by the data type.
    % 一张统计表格
    \autoref{Fig:analysis}(e) presents the frequently used data types and the preferred encoding channel. 
    
    Moreover, we have two interesting findings in our corpus.
    \add{First, among all elements in an image, some elements may contain additional semantic information, such as the two circles in the car referring to wheels.
    Designers prefer to encode such elements with correlated data.
    In a car, MPG (miles per gallon) is more relevant with the wheel than the car body, and designers prefer to use the wheel size to encode MPG.}
    % First, some elements in a metaphor may contain additional semantic information, such as the wheels in the car (\textbf{F1}).
    % Designers prefer to encode such elements with correlated data, like wheel size to encode MPG (miles per gallon)~\cite{surtola2005effect}.
    Second, data with similar attributes can be encoded in the same way.
    For instance, Chau et al.~\cite{chau2011visualizing} used different leaf elements in a metaphoric flower glyph to encode external and internal links of the webpage.
\end{itemize}

\subsection{Design Considerations}
\label{section: DCs}

We aim to design a system to generate MGVs automatically from a spreadsheet.
We summarize three primary DCs to guide the design.

\begin{description}[itemindent=-2.5em]
    \item[DC1\label{DC1}] 
    \textbf{Generate a semantically-resonant \mgv{}.}
    Using metaphors, successful \mgv{}s can promote the interpretation of the input data.
    Therefore, it is critical to choose an appropriate metaphoric design with respect to the data semantics.
    We likewise consider the rationality of the mapping between specific data dimensions and individual visual elements. 
    The system should consider both factors and ensure the quality of the final output \mgv. 
    % By integrating two considerations, the system will generate an \mgv{} which improves the data understanding.
    % % 强调两个步骤需要结合一起考虑
    % Therefore, the quality of the final output \mgv{} should be guaranteed by carefully selecting the metaphor and constructing an \mgv{} iteratively.

    \item[DC2\label{DC2}] 
    \textbf{Support automatic and efficient \mgv{} generation.} 
    Abundant online image resources provide design inspirations and create an opportunity for the automatic generation of \mgv. 
    % 自动
    However, selecting one appropriate metaphor image from abundant online sources is difficult.
    People must remember complex data features and consider multiple data mappings.
    Manual data mapping is labor-intensive since users need to calculate different attributes (i.e., size and angle) for encoding.
    Thus, we plan to automate the process to ease data exploration through quick generation, including image selection and data mapping.
    % Moreover, following \textbf{DC1}, we iterates the process in a loop without human effort.
    % 强调data-mapping 
    %For a better user experience, the final \mgv{} should be generated within a suitable period.
    %Due to plenty of pictures, there are various combinations for generating an \mgv{}.
    %Users may feel impatient waiting for a result if the recommendation comes from traversing all online images.
    %However, it is difficult to find an optimal solution in a limited time.
    %Therefore, we should balance the time and quality.
    %We set multiple filter conditions in different steps to produce a better \mgv{} in a short period.
    On the one hand, due to the large size of online images, it takes a long time to test various combinations of images to generate an appropriate result. 
    On the other hand, users feel tired of waiting for a result when all online images are required to be transformed and combined. 
    Thus, we set multiple filter conditions in different steps to balance the quality of \mgv{} and the time spent on the creation. 

    \item[DC3\label{DC3}] 
    \textbf{Integrate a mixed-initiative workflow.}
    Although an automatic system provides convenience, the generated visualization may not satisfy users' expectations. 
    Therefore, users should engage in the creation process~\cite{rubab2021Examining}.
    We consider a mixed-initiative workflow that integrates the machine's and human being's efforts.
    Our system provides some initial results for users to choose from and change by the given spreadsheet.
    Then, after modifying visual elements for specific data dimensions from users, MetaGlyph improves the final output and provides alternatives based on the users' preferences.
    To follow this practice, our system should provide a collaborative design workflow.
\end{description}

%% file: paper/4_Generation_Engine.tex
\begin{figure*}[htbp]
  \centering %图片居中
  \includegraphics{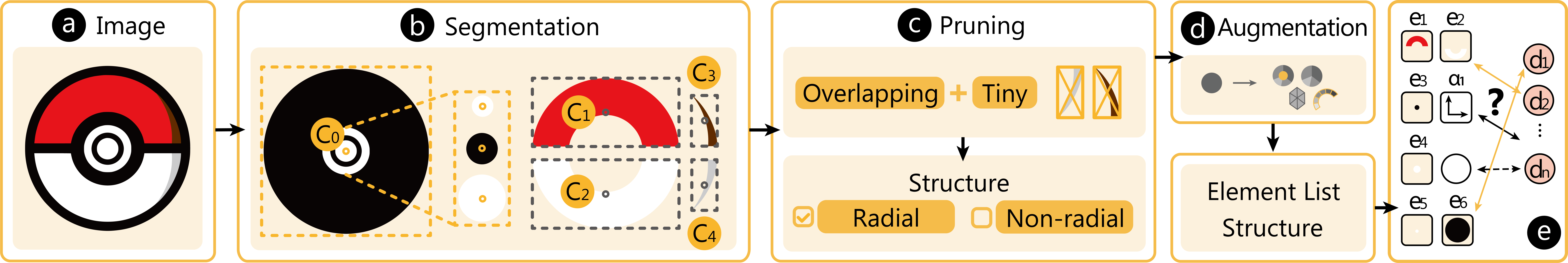}
  \caption{
    The entire process of metaphoric image selection for each 
    (a) input online image, including 
    (b) segmentation: dividing the image into visual elements, 
    (c) pruning: deleting redundant elements and determining the structure, and 
    (d) augmentation: checking if some elements can transform into charts.
    The output is an element list with an image structure.
    (e) The mapping space for constructing \mgv{}s.
    }
  \label{Fig.selection} %用于文内引用的标签
\end{figure*}

\section{MGV Generation Model}
\label{section: mgvGenerationModel}

This section introduces our two-step model to generate MGVs, including selecting metaphoric images and constructing \mgv{}s.

\subsection{Selecting Metaphoric Images}
\label{section:selection}
To construct valid \mgv{}s for the given input data, we first decide on appropriate metaphors with corresponding images. 
This subsection elaborates on the process of metaphoric image selection, including segmentation, pruning, and augmentation.
After obtaining candidates, we select one image each time for subsequent operations in \autoref{section: construct}.

% 图片最好符合glyph例子中的特征
Based on the findings from \autoref{section: analysis}, we summarize two criteria to determine the right image:
\begin{itemize}
    \item \textbf{C1:} It is semantically related to the data.
    \item \textbf{C2:} It is a vector image with a relatively simple structure.
\end{itemize}
First, we can derive \textbf{C1} directly given the definition of \mgv~\cite{kogan1980understanding}.
% as the definition we give in \autoref{section: analysis}.
% 直接找svg 然后用google搜索作为补充
For \textbf{C2}, vector images are easy to edit and reuse compared to raster images, which correspond to our needs for data binding.
% 简单
As discussed in \autoref{section: analysis}, it is more appropriate to determine the number of the encoded visual elements within the range from two to five.
Given that complex images will lead to high understanding costs, the source images of glyphs should be simple and easy to perceive.
According to \textbf{C1} and \textbf{C2}, we search eligible images on the Internet.

Given a spreadsheet, we can obtain general information about the dataset.
\add{The spreadsheet name is regarded as the topic of the dataset.
We choose \textit{SVG Repo}\footnote{https://www.svgrepo.com} to search for eligible vector graphics following \textbf{C1}.
The format of the vector graphics meets the condition, and their structure is simple for subsequent operations (\textbf{C2}).
In addition, we use the Google Image Search Engine as a supplement to ensure the diversity and abundance of candidate images.}
We use the keyword ``\textit{icon}'' and convert the resulting bitmap images into SVG using an open-source package \textit{Portace}.
As a result, we get a first-version candidate metaphoric image list in this step.
Next, we filter these candidates through subsequent processing steps, as shown in \autoref{Fig.selection}.

\textbf{Segmentation.}
We need to segment each image as elements from the previous output list.
% and determine the image structure.
Given that an SVG file comprises several paths, we convert each path into an individual SVG file to obtain an element list using the SVG format.
We derive $n$ files corresponding to the $n$ paths and calculate the center $C_i (i=1,2,...,n)$ of each retained element to determine the overall center of glyph $C_0$.

\textbf{Pruning.}
Given all visual elements, we prune non-essential elements following \textbf{C2} and determine the image structure.
% retain vital ones for data-binding.
% We remove two types of elements temporally: overlapping elements in a non-radial structured image and tiny elements for all structures, since hidden or tiny elements are not suitable for binding information.
% For images with a non-radial structure, elements which are completely overlapped conveyed similar information.
% For instance, many circles form a wheel.
% Compared with one circle, the advantage of such design mainly falls into beauty.
% Considering accuracy, we use the area of a specific shape rather than the bounding box to evaluate whether the element is tiny.
% We filter these elements for accurate data-binding.
To map data effectively, we remove tiny overlapped elements temporally, as associating the information with such elements will not improve the understandability.
\add{We heuristically found tiny elements were those occupying less than 0.5\% of the area of the whole image, which reached a good compromise between the understandability and aesthetics.}
% \add{Tiny elements are elements with a relatively small area.
% We heuristically found 0.5\% of the area of the whole image to be a good compromise between the forces.
% }
% Considering accuracy, we use the area of a specific shape rather than the bounding box to evaluate whether the element is tiny.
For instance, elements with center $C_3$ and $C_4$ in \autoref{Fig.selection} are removed in this step.
\add{Next, we decide on the image structure based on preserved elements.}
We follow the same classification as discussed in \textit{Glyph Layout} (\autoref{section: analysis}) since the image is the vital material to construct a metaphoric glyph.
The distinction is used for further data mapping.
We use the center position $\{C_0, C_1, ..., C_n\}$ to confirm the structure.
\add{As discussed in \autoref{section: analysis}, we check whether the origin of the polar coordinate system exists by analyzing all elements' center.
We first derive a possible rough origin position $P_o$ based on the center of the glyph.
If more than one element's center is close to $P_o$, we define this image as a radial structure and vice versa.
}
% If more than one element shares the same or close center with the glyph (), 
For example, in \autoref{Fig.selection}, since four circular elements and the whole glyph have the same center $C_0$, the image is regarded as a radial structure.
\add{For non-radial images, we check if the centers of all essential elements $C_i$ can be connected in a nearly straight line with slope $k$.}

\textbf{Augmentation.}
\add{In this step, we transform original basic shapes into charts and integrate them into the glyph design.}
According to our findings in qualitative analysis (\autoref{section: analysis}), we find 5 cases (star plot, donut chart, pie chart, heatmap, and boxplot) in existing research and design.
\add{Some charts are directly derived from a circular shape while others (heatmap and boxplot) need extra transformation~\cite{liu2017smartadp, LiuLZLWP16UncertaintyAware}.}
% All of them are derived from a circular shape.
Therefore, we add an extra tag for circular elements because they can transform into charts.

After the above operations, we get an element list.
We define the image as follows:
\begin{align}
  Image 
  &= \{ e_1, e_2, ..., e_n, S\} \notag\\
  &= \{ \{a_1, p_1\}, \{a_2, p_2\},..., \{a_n, p_n\}, S \} 
  \label{formula: elementList}
\end{align}
where $e_i$ refers to different visual elements, $a_i$ is a boolean variable indicating whether the element can be augmented, $p_i$ describes a path in an SVG file, and $S$ represents the image structure.
For a non-radial image, we also record the slope $k$.
For each image in the first-version candidates, we derive an element list for the following step.

\subsection{Constructing Metaphoric Glyph-based Visualization}
\label{section: construct}
This subsection introduces the method to construct an \mgv{}. 
We first formulate our problem into a mathematical form and then give a solution overview, including mapping data to different visual elements and placement as well as glyph rendering in MetaGlyph.

\subsubsection{Problem Formulation}
% In this part we formulate our problem and give an overview of the solution.

% 文字描述
To generate an MGV, we select several data dimensions from the given spreadsheet, pick some visual elements from the element list, map the selected data attributes to the chosen elements and determine the placement.
We also determine the visual encoding channel for each pair of one data attribute and an element.
The position attribute for glyph placement \add{(discussed in \autoref{section: analysis})} is also based on data.
% placement说明，只考虑xy坐标系+地图+类别
We adopt two variables, ${\alpha}_1$ and ${\alpha}_2$, to present both data-driven and structure-driven glyph placements.
The two variables are clear for data-driven placement to present numerical data due to a Cartesian coordinate system.
% Limited by a spreadsheet input, complex data structures like hierarchical ones are not suitable.
We consider several specific objects (map, timeline) and order relationships according to data features for structure-driven placement.
One variable ${\alpha}_1$ is enough to present this data group.
We need one categorical data dimension for the map and order relationship, and one temporal data for the timeline.
As the ultimate effect of the \mgv{} is determined by glyphs and placement, we consider them together.
% 是不是在sec3中要明确这个事情,整体的元素也有编码
Moreover, we take the whole image as an extra visual element $e_0$ because it often encodes data.

% 公式描述
We formulate the problem as:
given data $D = \{d_1, d_2, ..., d_n\}$,
% and an element-placement list 
% $L = \{e_0, e_1, ...,e_m, {\alpha}_1, {\alpha}_2, \varnothing\}$, 
we select $n$ mapping pairs $P$ for each data dimension to solve:
\begin{equation}
    \max \limits_{D, L} R_{mgv}(P_{D,L})
\end{equation}
where $R_{mgv}$ is the reward function for a solution,
$P_{D,L} = \{p_1, p_1, ..., p_n\}$ is all mapping pairs, as shown in  \autoref{Fig.selection}(e).
Each mapping pair can be represented as:
\begin{equation}
    p = d_{i} \leftrightarrow 
    \left\{
  \begin{array}{lr}
    e_{j} \\
    {\alpha}_{k} \\
    \varnothing
  \end{array}
  \right.
  \label{format: pairs}
\end{equation}
where $i \in [1, n]$, $j \in [0, m]$, $k \in \{1, 2\}$ and $\varnothing$ is an empty set.
$d_{i} \leftrightarrow e_{j}$ means the data dimension is mapped with an element (yellow arrows in \autoref{Fig.selection}(e)).
$d_{i} \leftrightarrow \alpha_{k}$ shows the data dimension is mapped with one axis of placement (the black arrow in \autoref{Fig.selection}(e)).
Since we do not present all data dimensions in an \mgv{}, some dimensions are not displayed.
$d_{i} \leftrightarrow \varnothing$ represents this condition (the black dotted arrow in \autoref{Fig.selection}(e)).

% 要求
Above all, we should guarantee the quality of the final \mgv{}.
The number of axes $\alpha_{k}$ is one or two to ensure a successful visualization.
For \autoref{format: pairs}, the amount of valid pairs ($d_{i} \leftrightarrow e_{j}$, $d_{i} \leftrightarrow \alpha_{k}$) is an unknown variable, which means we can choose one pair, two pairs, or even all data for data-mapping.
% 为什么要用MCTS
With the above considerations, the mapping space can be large for this question, even for a small dataset.
Moreover, intermediate results are not worth referring to until all mapping pairs are determined.
We cannot enumerate all solutions first and pick some reasonable ones as the output since a long waiting time for users violates \textbf{DC2}.
We address this problem using an efficient method named MCTS~\cite{montello2003testing}, which is proposed to search for the best next move in a game.
This algorithm can efficiently and logically explore a large space via a tree structure.

\subsubsection{Monte Carlo Tree Search}
\label{section: mcts}

\begin{figure}[tbp] 
  \centering %图片居中
  \includegraphics{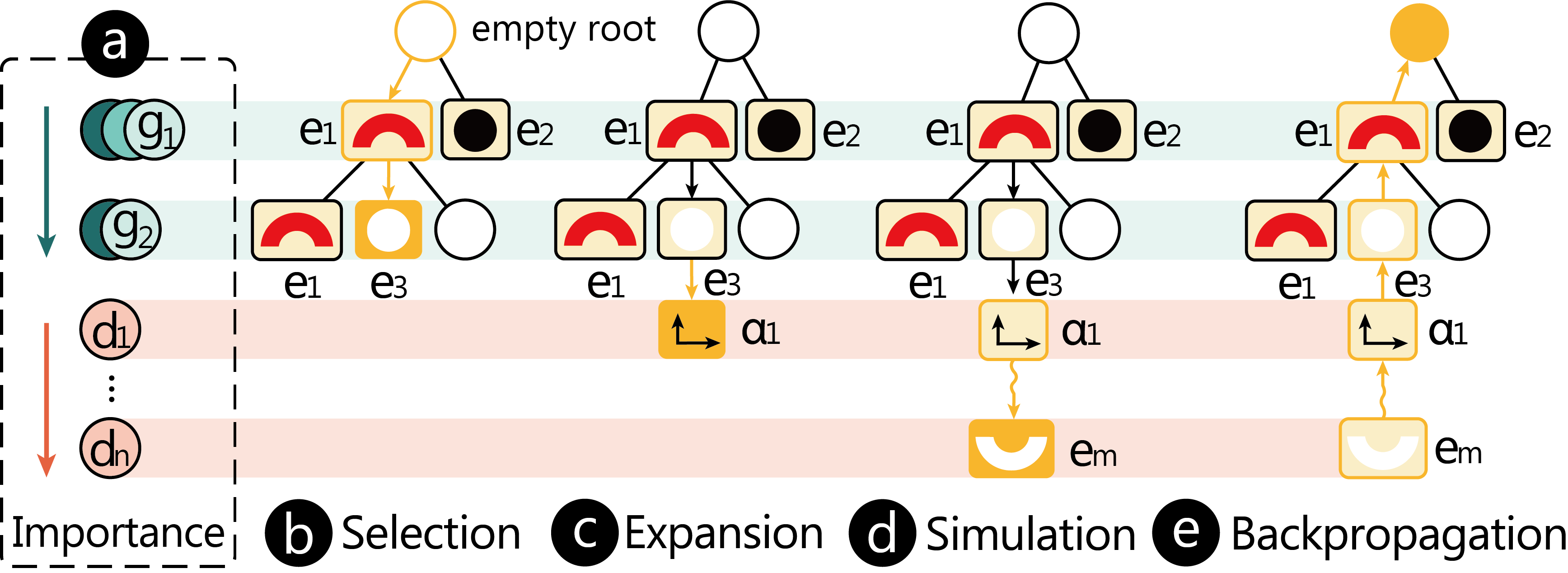}
  \caption{
    (a) The data dimensions and groups are ordered by importance independently.
    An iteration of the Monte Carlo tree search consists of four stages, including (b) selection, (c) expansion, (d) simulation, and (e) backpropagation.
    The shapes with a yellow frame represent the operation of this stage.
    Inside the nodes are element thumbnails ($e_i$) or axes ($\alpha_i$).
    Circles with no elements are empty nodes ($\varnothing$).
    }
  \label{Fig:mcts} %用于文内引用的标签
\end{figure}

We explore the mapping space by constructing a searching tree $T$.
Each tree node is a visual element $e_i$ or an axis $\alpha_k$ from the element-placement list. 
We start from an empty root node, as shown in \autoref{Fig:mcts}(b).
The mapping pairs (\autoref{format: pairs}) are represented as \textit{Node's Height} $\leftrightarrow$ \textit{Node}, where the \textit{Node's Height} corresponds to a data dimension $d_i$, and the \textit{Node} is one of three options in \autoref{format: pairs}. 
% The height of each node corresponds to a data dimension $d_i$ as shown in \autoref{Fig:mcts}(a), and the empty root node does not bind data.
In MCTS, nodes close to the root are usually explored more fully. 
Therefore, the order of data dimensions weighs a lot for the resulting mapping method.
We adopt an importance score to estimate the importance of all data dimensions and put the higher score dimension at the top of the tree.
% 对于group起来的数据，计算平均并优先考虑
% For similar data dimensions, we consider to use a same element  
The search process repeats four stages starting from an empty node: selection, expansion, simulation, and backpropagation.
Then, one mapping method is generated by a path from the root to a leaf node. 
After identifying one data dimension and an element, we choose an encoding channel according to the data type and element feature as discussed in \autoref{section:rendering}.
We also design a reward function to estimate the quality of all generated \mgv{}s.

% \textbf{Data Ranking.}
First, we need to order all data dimensions before building a mapping tree.
We estimate the importance of one data dimension based on its relevance to the data topic.
Semantic text similarity is commonly used in the natural language processing field~\cite{hu2019introductory}, like machine translation and image caption~\cite{li2021Image}.
Researchers use word embeddings to represent the original text information.
Since the description of one data dimension is usually a phrase aside from a word, we choose sentence embeddings.
Among existing state-of-the-art models~\cite{le2014distributed, mikolov2013efficient, reimers2019sentence}, we choose sentence-BERT~\cite{reimers2019sentence}, which is suitable for calculating sentence distance and avoids massive computational overhead.
We use cosine distance, which is widely used to measure distance in vector space.
Consequently, we rank the data dimensions by calculating the distances between the description of data dimensions and the data topic.

%For similar numerical data dimensions, such as the scores of different courses (i.e., math and music), we group them as $G_d$ because designers prefer to use the same encoding as discussed in \autoref{section: analysis}.
We group numerical data dimensions with similar meanings as $G_d$, such as math and music scores, and apply the same encoding method to the grouped  data dimensions.
While the visual elements are being mapped, these groups are treated the same as other data dimensions.
They cannot be considered as axes $\alpha$.
We compute the importance of the data dimensions by using the average distance:
$dist_G = \overline{dist_d}$, 
where $dist_d$ is the individual importance score of integrated data dimensions.
We also prioritize these groups and put them at the top of the tree ordered by their ranking as shown in \autoref{Fig:mcts}(a).

\textbf{Selection.}
A mapping tree is initialized with an empty root node, as shown in \autoref{Fig:mcts}(b).
In the selection stage, we aim to find the most promising element node $e_i$.
This step starts from the root node and selects a child node with the maximum upper confidence bound for trees (UCT)~\cite{browne2012survey} each time.
Generally, UCT considers the balance between less-visited and high-valued nodes:
\begin{equation}
    UCT = 
    \frac{r_i}{n_i} + c \sqrt{\ln {\frac{N_i}{n_i}}}
\end{equation}
where $r_i$ is the reward value, $n_i$ is the visited times of $e_i$, $N_i$ is the visited times of the parent node, $c$ is a constant.
After multiple experiments, we used $c=4$ since it performs better.
The selection stage ends until the most urgent expandable node is reached.
A node is expandable if it has unvisited (i.e., unexpanded) children.

\textbf{Expansion.}
One child node is added to expand the tree in the expansion stage.
We add a random visual element or axis node to expand as shown in \autoref{Fig:mcts}(c) and initialize the visited time and reward as zero.
Remarkably, we can add an empty node to expand because we do not present all data dimensions.
For individual data dimensions, all elements in the list $L = \{e_0, e_1, ...,e_m, {\alpha}_1, {\alpha}_2, \varnothing\}$ are alternatives.
However, for data groups, axes $\alpha_i$ are excluded.

\textbf{Simulation.}
A simulation process starts from a new node and uses a rollout policy to produce an outcome, as shown in \autoref{Fig:mcts}(d).
The rollout policy randomly chooses a node to expand recursively (like the step in \textit{the Expansion} stage) until the node cannot expand, which means all data dimensions have been mapped at that time.
To simulate more quickly, we do not pursue a high reward in this stage.
Our goal is to simulate more times to get a high reward in a limited time.
% This is the ``Monte Carlo'' part of the algorithm.

\textbf{Backpropagation.}
In this stage, the simulation result is backed up to update the selected nodes (\autoref{Fig:mcts}(e)).
New nodes are also added to the tree $T$ with a reward value and one visited time.
Other visited nodes need to update a reward value if it is bigger and add one visited time.
Then the search process returns to the selection stage or terminates when the time limits are exceeded, or the search tree is exhausted.

Finally, the path with the highest reward value is identified as the best matching method in this generating iteration.
% For this path, we define the selected visual elements as $E$ and the selected data dimensions as $D = \{D_d, D_g\}$, where $D_d$ is the list of individual dimensions and $D_g$ is the list of groups.

% \vspace{1pt}
\textbf{Reward Function.}
% If the number of axis nodes is not one or two, the reward is zero.
We propose a reward function to estimate the quality of the final \mgv{} via three criteria: importance ($I$), semantic relevance ($S$), and overlapping ($O$):
\add{
\begin{equation}
    R_{mgv} =
    \left\{
    \begin{array}{lr}
        \frac{1}{n}O_{mgv}\sum_{d,b \in \{e, \alpha\}}^{n}{I(d)S(d,b)}, 
            &\text{if } N_{\alpha} \in \{0, 1\}\\
        0,   &\text{otherwise} \\
    \end{array}
    \right.
    \label{equation: reward}
\end{equation}
where $e$ is one of the selected visual elements, $d$ is one data dimension or one data group, $\alpha$ is one of the axes, and $N_{\alpha}$ is the number of axis nodes.}
We should check the number of axes ($\alpha$) in the derived mapping method
to guarantee a valid \mgv{}.
We removed all empty nodes at that time due to their zero importance.

\begin{itemize}
    \item \ul{\textit{Importance Score}} estimates the importance of one data dimension or group.
    Since a cosine distance orders the importance mentioned above, some scores will be negative, which violates the reward calculation of MCTS.
    We normalize the distance as \textit{Importance Score} $I$ to ensure the feasibility of our model.
    
    \item \ul{\textit{Semantic Score}} estimates the semantic relevance between a visual element and one data dimension.
    As discussed in \autoref{section: analysis}, designers prefer to encode visual elements with correlated data.
    We adopt the Transformer-MM~\cite{chefer2021generic} to bridge a text and image.
    Chefer et al.~\cite{chefer2021generic} used the attention layers of the model to produce relevancy maps for image and text interactions.
    We choose the model CLIP~\cite{radford2021learning} to derive our semantic relevance score due to its abundance.
    It learns from 400 million text-image pairs already publicly available on the Internet.
    Specifically, we first convert the origin SVG into a pixel image. 
    Given the pixel image and a textual description of one data dimension ($d \in D$), we derive a heatmap of pixels corresponding to the description and normalize its value.
    Next, for a visual element $e$, we calculate the average relevance of the area covered by it as the final $S$.
    We adopt different scores based on data type separately for axes.
    When presenting temporal data or geospatial data, the $S$ is assigned one since such placement is semantically-resonant (\textbf{DC1}).
    We use the entire image for other data types to derive the relevance score $S$ because the placement is relevant to all elements.
    
    \item \ul{\textit{Overlapping Score}} estimates the overlapping degree of the final \mgv{}.
    We calculate the overlapping areas in the \mgv{} rendered in \autoref{section:rendering} based on the bounding box of each glyph:
    \begin{equation}
    O_{mgv} =
        \left\{
        \begin{array}{lr}
            1  &\text{if } P_{overlap} \leq 30\%\\
            0  &\text{otherwise} \\
        \end{array}
        \right.
    \end{equation}
    where $P_{overlap}$ is all elements' average overlapping percentage.
\end{itemize}

% Definition
% MCTS 每个步骤
%   Initialization/Selection/Expansion/Simulation/Back propagation
%   Reward function
    % Semantic / Importance

\subsubsection{Rendering}
\label{section:rendering}

Given a visual element, the glyph placement, and data dimension, we need to make the encoding channel clear for rendering.

\textbf{Encoding Channel for Visual Elements.}
As mentioned above, all data dimensions are divided into $d$ (one data dimension) and $g$ (a data group).
Moreover, we divided all visual elements into $e_a$ and $e_{!a}$ for different value of the augmentation tag $a_i$ retained in \autoref{section:selection}.
$e_a$ corresponds to $a=1$ and vice versa.
We consider three conditions separately for different visual elements and data dimensions:
\begin{itemize}
    \item \ul{$d \leftrightarrow \{e_a, e_{!a}\} $.}
    Since the data input is only one dimension, we treated $\{e_a, e_{!a}\}$ the same.
    Although element $e_a$ can be augmented, we preserve its original shape for a better representation.
    We use a heuristic method based on examples in our corpus (\autoref{Fig:analysis}(e)).
    We determine the encoding channel by analyzing the shape of elements and the data type.
    \add{Given a data dimension, we select the most frequent encoding channel based on its type.
    When two or more data dimensions share the same element, we order them based on the reward function (\autoref{equation: reward}).
    Dimensions with a higher reward can occupy a more frequent encoding channel.}
    When using size channel, we need an additional decision based on the image structure (\autoref{formula: elementList}).
    Height, length, and area are three options for ultimate encoding.
    
    \item \ul{$g \leftrightarrow e_a $.}
    For the pair of data groups and elements that can augment, we transform the element into different charts to present all dimensions in the group.
    We adopt four charts, namely, pie charts, donut charts, star plots, and heatmaps as shown in \autoref{Fig.selection}(d).
    % 3.2中要说明三种数据类型
    For numerical data, we consider star plots.
    Specifically, pie charts, donut charts, and heatmaps are alternatives for proportional data.
    
    \item \ul{$g \leftrightarrow e_{!a} $.}
    We use a typical design in our corpus with visual elements that cannot augment.
    We choose three encoding channels (rotation, color, and size) of one visual element to represent a data group.
    Rotations and colors are both used to distinguish different data dimensions in the group.
    The size encodes the numerical data.
    An example is displayed in \autoref{fig:teaser}(b).
    The leaves are rotated to illustrate the forest area in different years.
    Color is also utilized to distinguish the value year, as shown in the legend (\autoref{fig:teaser}(b1)).
    The size of the leaves represents the percentage value.
\end{itemize}
Notably, we check if the removed elements $e_r$ in \textit{Pruning} (\autoref{Fig.selection}(c)) need to be drawn after confirming encoding channels.
We call the preserved element that overlaps with the removed one as $e_p$, and the encoding channel for $e_p$ as $c_p$.
If $e_p$ encodes data and $c_p$ is size, we scale the $e_r$ the same as $e_p$ and add it into the glyph.
If $e_p$ is transformed into a chart or does not encode data, we delete the element.
For other conditions, we keep it in its original shape for drawing. 

\textbf{Glyph Placement.}
Owing to the limit in the simulation stage of MCTS, we only need to consider the independent data dimensions $d$ and the axes $\alpha$.
For temporal data, we use a timeline to illustrate following \textbf{DC1}.
For geo-spatial data, we utilize a map (\autoref{fig:teaser}(b) and (f)).
For other numerical data, if the number of axes is one, we adopt a horizontal axis to place glyphs.
A Cartesian coordinate system is the right choice with two axes, as shown in \autoref{fig:teaser}(a) and (e).
Concerning categorical data, we place glyphs in a specific order as shown in \autoref{fig:teaser}(c) and (g).

% 所有的图片
Ultimately, we can obtain one semantically-resonant \mgv{} with the highest reward for each image candidate.
We pick the greatest as the output based on all candidates' reward values, and others are ranked as alternatives for users to choose from.

%% file: paper/5_MetaGlyph.tex
\section{MetaGlyph}

\begin{figure*}[htbp] 
  \centering %图片居中
  \includegraphics{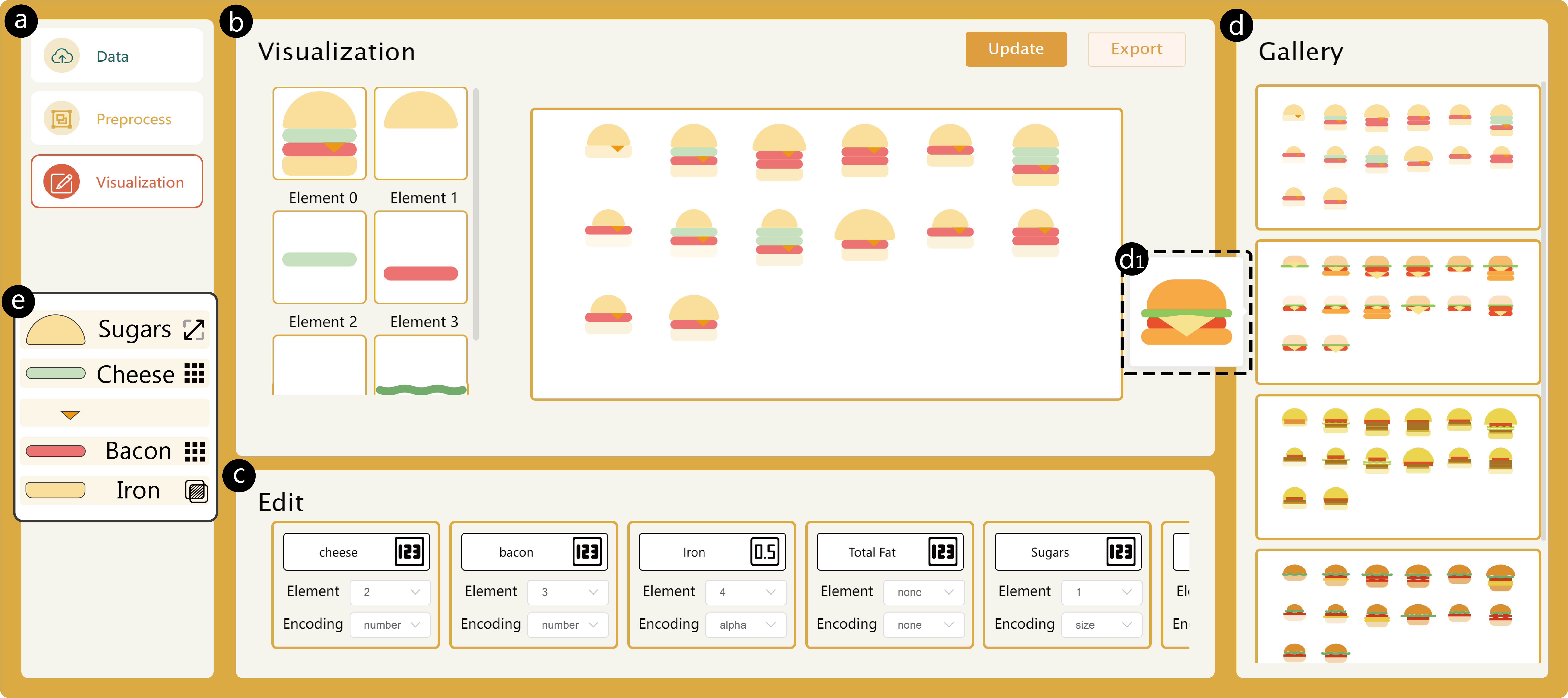}
  \caption{
  The interface of MetaGlyph is consisted of :
    (a) a Menu view for different steps;
    (b) an \mgv{} view representing visual elements and corresponding \mgv{}, 
    (c) an Edit view, in which the data mappings can be modified based on user's preferences, and
    (d) a Gallery view with other \mgv{} options.
    (e) concrete data mappings for the \mgv{} view in detail.
    % The legend button (e) can be clicked to represent data mappings for the \mgv{} view in detail.
    }
  \label{Fig:metaglyph} %用于文内引用的标签
\end{figure*}

This section introduces the workflow of MetaGlyph that generates \mgv{}s given a spreadsheet input and the interface design.
% After that, we propose the interface design. 

\subsection{Workflow}
We design the workflow of MetaGlyph based on the design considerations proposed in \autoref{section: DCs}.
Following \textbf{DC3}, MetaGlyph integrates the automatic model and user interactions into the authoring process.
First, users need to import a spreadsheet of data as the input. 
An initial \mgv{} is generated based on the automatic model.
The quality of \mgv{} is ensured by a reward function that considers different dimensions, including data importance and semantic matching (\textbf{DC2}) as discussed in \autoref{section: mgvGenerationModel}.
% The automatic model consists of two steps: (1) metaphoric image selection and (2) \mgv{} construction.
% Based on a search engine, the former uses text information to look for an appropriate image and divide it into different elements.
% The latter is designed based on the Monte Carlo Tree Search (MCTS) to get a suitable mapping for the data dimensions given by the input spreadsheet and the elements from the image search.
% Following \textbf{DC1}, the above two steps are implemented iteratively to provide a sound output.
After that, users can modify the data mappings and the encodings of the initial visualization result.
The automatic model will re-calculate and create a new \mgv{} based on users' input.
To ensure the efficiency of the computing following \textbf{DC2}, we limit the search number of images at one time.
Next, users can smoothly switch between different visualizations and refine the satisfactory one as the final output.
Given that both the users and the model contribute to the design of the \mgv{}, MetaGlyph can derive more novel and creative designs compared with a solo-authoring workflow.

\subsection{User Interface}
The MetaGlyph (\autoref{Fig:metaglyph}) consists of four views: Menu view, \mgv{} view, Gallery view, and Edit view.
The Menu view shows the creation procedures of our system, that is, first upload, then preprocess, and ultimately edit the visualization.
% 上传+数据呈现
Our system only needs one data input.
Users can upload a spreadsheet, and our system will process the data.
% and then group several data columns in the same type for their needs in the proprocessing step as shown in \autoref{Fig:metaglyph}().
% These data groups are regarded the same as other data dimensions.
% Then our system will process the data.
% mgv
After the system's calculation, we can derive several \mgv{}s with high rewards within a suitable waiting time (\textbf{DC2}) using the \mgv{} generation model in \autoref{section: mgvGenerationModel}.
Moving to the edit step, the \mgv{} with the highest reward is presented in the \mgv{} view (\autoref{Fig:metaglyph}(b)), and the other alternatives are shown in the Gallery view  (\autoref{Fig:metaglyph}(d)).
Moreover, all segmented visual elements of the image are displayed on the left of the \mgv{} view.
Here, \textit{Element 0} represents the entire metaphor.
% alternatives
Users can select other \mgv{}s in the Gallery view following their preferences.
When hovering over the alternatives, we display the original metaphor (\autoref{Fig:metaglyph}(d1)).
% edit
The Edit view will show all mappings and allow users to modify them (\autoref{Fig:metaglyph}(c)).
Each small panel illustrates detailed information about one data dimension or one data group, including the title of the data column, data type, element mapped, and corresponding encoding channel.
The ordering of different data dimensions depends on the importance score discussed in \autoref{section: construct}.
We use a small icon to illustrate the data type after its title.
For data that are not represented in the current \mgv{}, the mapped element is shown as \textit{None}.
Users can alter the data mappings by changing the visual elements in corresponding panels in the Edit view.
After clicking the \textit{Update} button in the \mgv{} view, the generation model will re-calculate to obtain a satisfactory result following \textbf{DC3}.
% Minor changes are also supported after ensuring the data mapping.
% Users can alter some attributes for different encoding channels like the colors and sizes.
Finally, by clicking the \textit{Export} button, users can export the \mgv{} shown in the center as an SVG file.

%% file: paper/6_Evaluation.tex
\subsection{Usage Scenario}
\label{section:scenarios}
We present one usage scenario in detail in this section.
% Following a similar procedure, we demonstrate the expressiveness of MetaGlyph by creating more cases as shown in \autoref{fig:teaser}.
% 
% \textbf{Scenario 1: }
% The first scenario demonstrates the entire process of generating an \mgv{} using MetaGlyph.
Adam is a junior visualization researcher.
\add{His task was to analyze the ingredients and nutrients of different hamburgers in McDonald~\cite{kaggleMcDonald} for customers to make a better choice.}
He uploaded the processed tabular data in the MetaGlyph system and clicked the ``generate'' button.
After a few seconds, the system generated a list of \mgv{}s, as shown in \autoref{Fig:metaglyph}.
The \mgv{} in the center represents three data dimensions: burger lettuce, sugars, and bacon (\autoref{Fig:metaglyph}(e)).
After browsing and absorbing the concrete mapping relations displayed in the \textit{Edit} view (\autoref{Fig:metaglyph}(c)), Adam quickly understood the encodings of the glyph.
Bread is a kind of carbohydrate.
Therefore, the size of the upper bread is appropriate to encode the data attribute \textit{Sugars}.
He was also satisfied with the method to show \textit{Burger Lettuce} and \textit{Bacon} using green and red lines (\autoref{Fig:metaglyph}(e)) due to the intuitiveness.
He clicked three titles of corresponding data dimensions in the \textit{Edit} view (\autoref{Fig:metaglyph}(c1)) to add three limits for the next update.
Next, Adam obtained a new result and continued exploration for other alternatives in the Gallery view (\autoref{Fig:metaglyph}(d)).

\begin{figure}[!ht] 
  \centering %图片居中
  \includegraphics{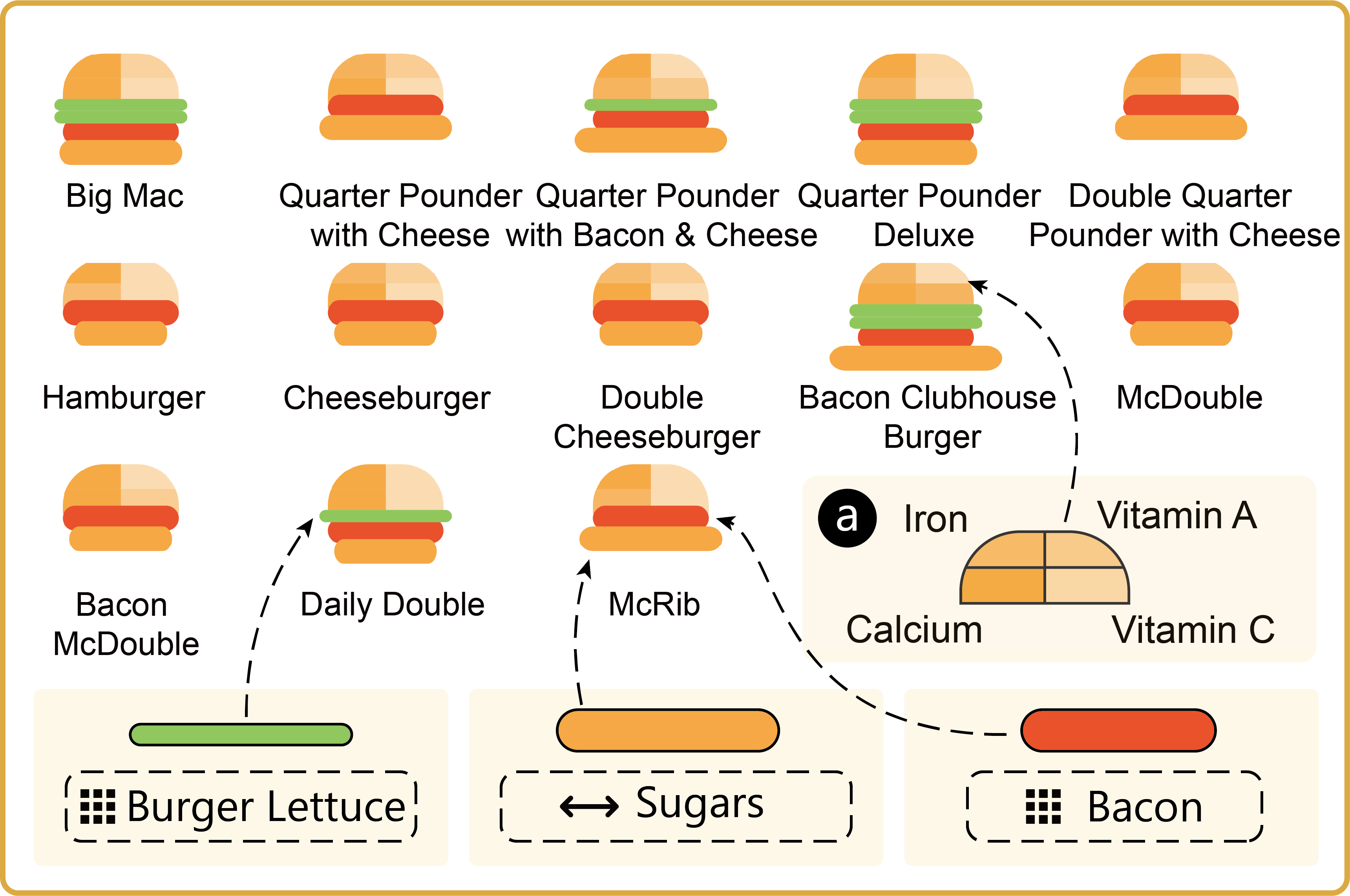}
  \caption{
  An example generated with the burger dataset.
  (a) The legend for the upper bread generated by MetaGlyph.
  The boxes with yellow backgrounds illustrate the data mappings and encoding channels.
  Dashed arrows are used to associate visual elements with the \mgv.
  }
  \label{Fig:example} %用于文内引用的标签
\end{figure}

Reanalyzing the dataset, he considered grouping some data columns, including iron, calcium, vitamin A, and vitamin C, since these are often considered as nutrients.
% often appear together on food products.
He returned to the \textit{Preprocess} step (\autoref{Fig:metaglyph}(a)) and added one data group.
This time he found an interesting \mgv{}, which adopted a heatmap-like design on the upper bread to represent the new data group.
Adam preserved this design and adjusted other mappings following his preferences.
After clicking the legend button (\autoref{Fig:metaglyph}(b1)), he understood the encodings for the upper bread (\autoref{Fig:example}(a)).
Using such a design, he could grasp different levels of four components at first sight.
Other encodings are displayed in \autoref{Fig:example}.
The numbers of red lines and green lines illustrate the amount of bacon and lettuce layers, respectively.
The length of the bread below represents the \textit{Sugars}.
Satisfied with this result, he exported this \mgv{}.
% 根据最后这次demo生成的结果

\add{Following a similar procedure, we demonstrate the expressiveness of MetaGlyph by creating more MGVs with various datasets ( \autoref{fig:teaser}).
These examples are hard to create with existing tools. 
For example, automatic tools (e.g., GlyphCreator~\cite{Ying2022GlyphCreator}, Diatoms~\cite{ Brehmer2022Diatoms} do not provide explicit support for metaphoric glyph design. 
Authoring tools (e.g., DDG~\cite{Kim2017DDG}, DataQulit~\cite{ZhangSBC20DataQuilt}) provide greater flexibility and allow users to achieve similar results.
% we derive more use cases to demonstrate that MetaGlyph can create MGVs for various datasets as shown in \autoref{fig:teaser}.
% Some tools can obtain a similar result, such as several manual tools in visualization (e.g., DDG~\cite{Kim2017DDG}, DataQulit~\cite{ZhangSBC20DataQuilt}).
However, the authoring process is tedious and time-consuming (e.g., requiring users to find or draw suitable metaphor images and encode data), and the quality of MGVs is highly dependent on the user.
Moreover, they do not support embedding charts in the glyphs (e.g., star plots in \autoref{fig:teaser}(a) and pie charts in \autoref{fig:teaser}(f)).
% Graphics software (e.g., Adobe Illustrator) is able to generate all example in \autoref{fig:teaser}.
% Users can draw everything using such tools, including different charts.
% The price of freedom is a lot of calculation for elements' parameters (e.g., size, color), which is repetitive and error-prone.
}

\section{Expert Interview}
\label{sec:interview}
% This section presents the expert interviews to evaluate the system.

% \subsection{Expert Interview}

To evaluate the MetaGlyph system and generated \mgv{}s, we performed a series of interviews with three expert users in different fields.
The first expert (E1) is a data journalist who has worked for a digital-news firm for more than three years.
The second expert (E2) majored in visual communication and has more than ten years of design experience.
The third expert (E3) is a senior researcher who has studied human-computer interaction and data visualization for more than eight years.

\textbf{Procedure.}
We conducted the interview via an online meeting system.
Each interview began with a 10-minute introduction of \mgv{} following a 10-minute demo of our MetaGlyph system.
% 是否使用？
Next, the experts were encouraged to use MetaGlyph online on their own.
After thorough trials of our system's features, we asked experts to generate \mgv{}s using the \textit{Burger} and \textit{Pokemon} datasets introduced in \autoref{section:scenarios}.
We recorded their creation process in a think-aloud approach. 
% Interviewees were encouraged to edit the initial result and discuss their considerations.
Our interviews were seeded with a set of questions that probed into the effectiveness of MetaGlyph and the quality of the generated \mgv{}s.
The one-hour screen capture videos and audios were recorded from these interviews for further analysis.

\textbf{Feedback.} 
In general, all experts expressed their compliments of MetaGlyph and agreed on its promising usage.
We summarized the interview results from three perspectives:

\ul{Workflow.}
Three interviewees appreciated the overall design of the workflow.
\add{All of them expressed the effectiveness of integrating metaphors into the design.
They believed that the automatic system is an efficient way to speed up the generation.}
During their previous creation, they preferred design software (e.g., Adobe Illustrator) or programming (E1, E3) after confirming the design ideas.
E1 said, \quo{Some of my colleagues write codes, but it requires long-time learning.}
E3 commented, \quo{When using design software, I need to do time-consuming batch work.}
She also expressed the difficulty of dealing with some visual elements, such as adjustments of arc angles.
With MetaGlyph, she thought the creation process would be more straightforward.
% He further suggested that color can also be considered in understandability.

\ul{System.}
All experts were satisfied with the design of MetaGlyph.
They appreciated the aesthetics of the interface and expressed the ease of learning cost when using MetaGlyph.
\add{E1 noted, \quo{The interaction is intuitive, and I can use it proficiently after a simple demonstration.}
For the designs of different views, E2 observed the \textit{Gallery view} and commented, \quo{Due to the difference in users' visual experience, their understanding ability is disparate. It is considerate to provide various alternatives for them.}
E3 liked the \textit{Edit} view, \quo{I prefer to try different results on my own. Although the system's results are nice, I can derive a different version by changing some attributes in the edit view.}}
% E3 gave some advice on MetaGlyph from the perspective of an interaction designer.
% She suggested exporting more files (e.g., legend, data mapping information, and visual elements) together with the \mgv{}, allowing further refinement using professional design tools.
% % She also made suggestions to improve the system interface, such as changing buttons and highlight styles.
% We further revised the system according to her suggestions.

\ul{Visualization.}
In terms of the quality of generated \mgv{}s, all interviewees agreed that the outputs were thoughtful due to the semantic relevance and metaphor embodied.
E2 underscored the match pattern for data and visual elements in MetaGlyph, which he also focused on during his design.
He said, \quo{Important data should be mapped with prominent elements.}
He also agreed with the limitation of the number of encoded data dimensions, \quo{It is essential to make trade-offs to data, such as preserving three data dimensions. Too much information will lead to comprehension problems.}
E1 and E3 both commented that \mgv{}s could serve as the basis for creative work in their early design.
E3 said, \quo{When designing, the most struggling part is the beginning due to the large design space. 
\add{The outputs not only give me a direction but also broaden my scope via some unexpected designs.}}
She agreed to use online sources since such images meet most people's cognition.
Therefore, the \mgv{}s can be better understood by audiences.
% Experts also gave suggestions on the resulting \mgv{}.
% For example, except for mapping information in \textit{Edit} view, E2 suggested adding more annotations and textual descriptions into the \mgv{}s.
% E3 thought it would be better to augment the visual appearance of visual elements if they were being used, as discussed in \autoref{section: construct}.

\ul{Suggestions.}
We also received suggestions from different aspects.
As an editor, E1 focused on the stories behind the data and suggested that corresponding conclusions can be generated automatically together with \mgv{}s.
\add{E2's advice fell into the improvement of visual comprehension.
He was confused about some specific values in the output MGV such as the size of the burger layer represents.
E2 suggested adding more annotations and textual descriptions into the \mgv{}s.}
He noted that color could also be considered in understandability.
E3 gave some advice from the perspective of an interaction designer, such as the button position and the highlight effect.
% She thought it would be better to augment the visual appearance of visual elements if they were being encoded, as discussed in \autoref{section: construct}.
She also suggested exporting more files (e.g., legend, files of different elements, mapping space between data and elements) together with the \mgv{}, thereby allowing further refinement using professional design tools.
We further revised the system according to her suggestions.

%% file: paper/7_Discussion.tex
\section{Discussion}

This section reflects the implications and limitations of MetaGlyph.

% \subsection{Implications for Generating \mgv{}s}
\subsection{Implications}
In this paper, we contribute an automatic approach to generating glyph-based visualization associated with comprehensive visual metaphors to help people understand data in an intuitive manner~\cite{Fuchs2017GlyphReview}. 
We propose the semantic-based generation framework by selecting resonant metaphorical images and mapping visual elements with data dimensions considering data importance and semantic relevance.
Users can customize the resulting \mgv{}s based on their requirements. 
We derive a set of implications in the design process of our system. 

\noindent \textbf{Automating glyph designs in association with semantic-resonant metaphors.}
Metaphors are widely used in visualization, which associates  data semantics with figurative motifs.
MetaGlyph aims to automatically generate glyphs whose visual elements are associated with semantically related metaphors.
To this end, we first search for appropriate metaphors with images considering the overall topic of the input data. 
Furthermore, we decompose visual elements of the metaphor and match them with data attributes with respect to the underlying semantic relations by adopting a prediction model. 
For example, the system searches burger images for the dataset of burger ingredients and maps the lettuce and bacon to the green and red elements, respectively (\autoref{Fig:example}).
In this way, the resulting \mgv{}s are associated with the underlying data semantics cohesively. 

\noindent \textbf{Balancing design expressiveness and perceptual effectiveness.}
The increase in data dimensionality leads to a complicated metaphoric glyph design that is difficult to understand~\cite{Borgo2013GlyphbasedVisualization}.
Our qualitative analysis (\autoref{section: analysis}) and expert interview (\autoref{sec:interview}) both indicate the constraint of the number of encoded data dimensions. 
Besides, our expert E2 suggests emphasizing important data with prominent visual elements,
\quo{I will not utilize all elements for presenting data. It would be overwhelming.}
Some visual elements in the metaphor are better used to encode data attributes, while others are more suitable for decoration purposes.
Accordingly, the encoded data dimensions and visual elements should be balanced with perceptual constraints.
Our current approach has filtered \mgv{}s that utilize overmuch elements in the mapping space (\autoref{section: mcts}). 
Further experiments are encouraged to explore the appropriate amount of conveyed information in an \mgv{} that informs better automatic generation criteria with perceptual effectiveness.

\noindent \textbf{Supporting a human-machine teaming approach.}
We situate MetaGlyph as a human-machine teaming system that integrates the machine with users' feedback.
Incorporating metaphors into designs is generally considered a highly creative task, requiring implicit knowledge to determine the right imagery and modify it to fit scenarios. 
Faced with the large search space of metaphors and data mappings, generating \mgv{}s is time-consuming and laborious.
Automating the labor-intensive parts with machines is highly required.
Moreover, MetaGlyph provides more than tool-level assistance.
%It explores horizontally within a given search space to help users explore different initial designs.
It explores the search space for users to compare design alternatives.
Both novices and experts can benefit from the rapid results of the initial designs.
Users are likewise supported to modify the results based on their preferences.
With the machine's assistance, users can focus on the data mappings without adjusting the elements' attributes manually.
Moreover, \mgv{}s can be updated multiple times according to users' requirements and the machine's re-calculation.
By assigning the labor-intensive part to the machine and providing the connectors of subjective decisions to human beings, MetaGlyph provides a successful human-machine teaming example and inspires the design of visualization tools in the future.

% \subsection{Implications for Resulting \mgv{}s}

% \textbf{Balancing the Encoded Data and Understandability.}
% The goal of integrating metaphors into glyphs is to help people understand data promptly and accurately~\cite{Fuchs2017GlyphReview}.
% However, the increase of data dimensionality will inevitably lead to a complicated metaphoric glyph design which is difficult to understand~\cite{Borgo2013GlyphbasedVisualization}.
% As E2 said, \quo{some elements are more suitable as decoration, while others are better represented to encode data. I will not utilize all elements for data binding.}
% We also adopt this idea and balance the trade-offs between semantic associations and information to communicate.
% As discussed in \autoref{section: mcts}, we filtered \mgv{}s that map overmuch elements in the mapping space.
% Moreover, E2 noted that \quo{Transforming abstract data into graphics needs thought and refinement for human designers.}
% Those comments call for continued research on more accurate data expression using \mgv{}s, such as utilizing more than metaphors and multiple elements for one data dimension.

\noindent \textbf{Providing design inspirations.}
% 强调空间很大 人只会尝试习惯的
% 趋向于更好
All experts appreciated how MetaGlyph allowed them to discover unexpected designs. 
Given that the design space of \mgv{}s can be much large considering both metaphor selection and visual mapping, it would be impossible for designers to try all alternatives.
In addition, as mentioned by E3, designers are likely to start with  familiar designs based on their usual practice.
Instead, the machine enumerates all possibilities, which may result serendipitously in promising designs.
For example, E3 praised the star plot inside the Pokeball (\autoref{fig:teaser}(a)) and the flower-like leaves (\autoref{fig:teaser}(b)).
\quo{I never thought of using such designs for these datasets. The visualization created by MetaGlyph shows diverse design possibilities,} E3 said.
% \sxh{@yinglu: add an example showing that one expert may mention which design he/she never though of and inspires further design?}
Designers can further refine the design using MetaGlyph or export the whole design assets for fine-tuning with professional tools.

\add{\subsection{Failure Cases and Limitations}
\textbf{Failure Cases. }
We observed several wrong or bad cases during the development of MetaGlyph. 
First, one critical source of some failure cases is the diversity of images the system searches from online sources as the metaphors.
For example, some images may have elements in multiple layers, such as \autoref{Fig.fail}(a) shows a car with a background. 
% a car image with background (\autoref{Fig.fail}(a)).
% Since the background does not entirely cover the car, we preserve both elements in \textit{Pruning}.
% However, irrelevant with the car dataset, the background does not encode data.
% Therefore, it still appears in the final \mgv{} as a component of the metaphor.
When scaling the car to encode data, the system cannot guarantee that the car can always be within the background, resulting in an incomprehensible \mgv{}.
Second, as we currently adopt some heuristic methods to decide the layout (\autoref{section: construct}), it cannot always be true. 
For example, \autoref{Fig.fail}(b) is wrongly identified as a radial glyph as it has two elements whose centers are close to the SVG's center. 
% some heuristic methods may result in wrong judgments when selecting metaphoric images (\autoref{section: construct}).
% For example, in some conditions, a non-radial glyph may also have two elements whose centers are close to the SVG's center (\autoref{Fig.fail}(c)).
Future work can use new techniques in computer vision and deep learning to optimize our framework.
% heuristic -> 智能

% We utilized online images as metaphors.
\begin{figure}
  \includegraphics{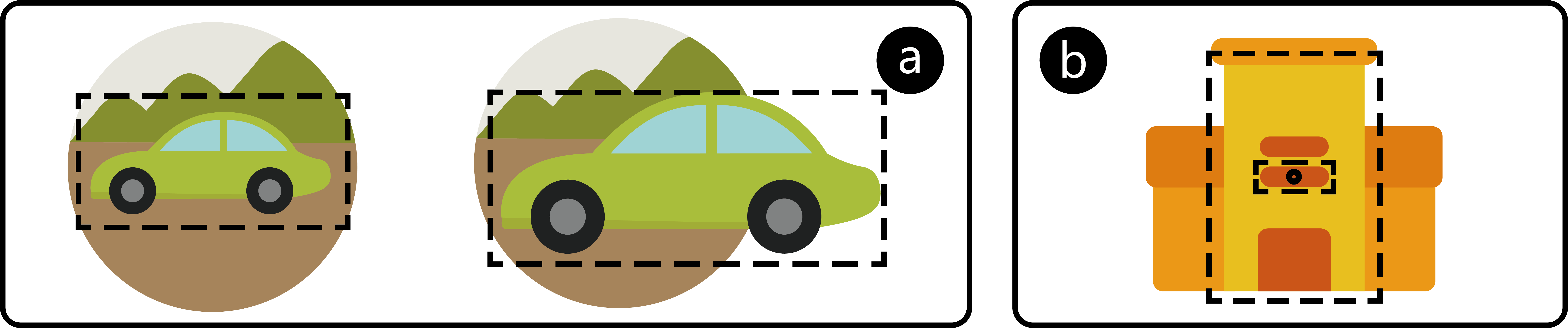}
  \caption{
  (a) A car with a driving background.
  (b) A hotel.} %最终文档中希望显示的图片标题
  \label{Fig.fail} %用于文内引用的标签
\end{figure}

\textbf{Limitations. }
% 更加定制化，可以修改一些边界值/风格/重要数据
First, the degree of customization for MetaGlyph is limited.
Specially, we follow the default style of the metaphoric source image, and use fixed parameters when calculating (e.g., set the boundary $0.5\%$ in Pruning in \autoref{section:selection}).
Future work can allow users to customize their MGVs, including but not limited to styles~\cite{chen2022Review} and a custom field of parameters.}
% First, we follow the default style of the metaphoric image in generating \mgv{}s.
% However, styles can provide extra semantic information, e.g., using bright colors to express positive emotions.
% That is, the style in the generated \mgv{}s may not be always consistent with the input data semantics. 
% Future work can optimize the style when designing \mgv{}s. 
% Thus, we plan to add style mapping to improve the semantic relevance; in turn, the comprehensibility of an \mgv{} will be improved.
% input单一
Second, MetaGlyph only supports tabular data with a single sheet, while complicated data structures, such as graph-related data, are not supported.
Because graph-related data can be represented as multiple-sheet data (e.g., adjacency matrices), we plan to extend MetaGlyph with more data processing operators to integrate multiple data sheets in the encodings of a single glyph.
In this way, we can improve the diversity of the resulting visualization for MetaGlyph.
%Aimed at generating glyph-based visualization, the data users want to present can be complex.
% A possible solution is to convert complex data into several data tables. 
% The system can be extended to generate \mgv{}s with diverse glyph placements.
% r2 Another possible use for the proposed technique is in pedagogical settings, such as when teaching data literacy or visual encoding; however, this possibility goes unmentioned. 
% data mapping 考虑不够全面 重要的元素 圆形结构在中间的/大小和重要性的关系/不同encoding之间的冲突问题
\add{Third, we mainly consider the importance of data dimensions and their semantic relevance with visual elements in data mapping. 
Other factors (such as SVG elements' sizes and shapes, mapping different elements with different channels, and users' priori knowledge) also play roles in the mapping. 
Future research can further explore how to consider all these factors to achieve a better MGV design.
% Third, we do not consider thoroughly in data mappings since we focus on metaphors and semantic relevance.
% The relation between position and importance, size and importance need to be taken into consideration in the future.
% We also plan to add more detailed rules to solve the conflicting problems caused by different encoding channels (e.g., one element uses size and another uses color).
}

\section{Conclusion and Future Work}

This paper presents MetaGlyph, an automatic system for generating \mgv{}s.
We first conduct qualitative analysis to understand the design patterns of \mgv{}s and then develop an automatic generation framework.
We automate the entire process by searching for appropriate metaphors, processing images, and determining multiple data mappings along with visual encodings.
Given the ample mapping space between visual elements and data dimensions, we adopt an MCTS algorithm to achieve an efficient and effective search considering data importance, semantic relevance, and glyph separation.
By uploading tabular data, our system will generate several \mgv{}s within a few seconds.
Moreover, as a human-machine teaming system, MetaGlyph enables users to modify the \mgv{}s following their preferences.
We evaluate MetaGlyph through two usage scenarios and a gallery of examples and demonstrate its effectiveness via a series of expert interviews.
% We plan to consider style customization and diverse data inputs to support more creative designs.
\add{MetaGlyph shows its potential in various situations, such as education~\cite{kui2022Survey} and journalism.
Teachers can use MetaGlyph to create metaphoric designs to help teach data literacy or visual encodings, while journalists can use it to create figurative and expressive visualization easily and quickly. 
% For teachers, when teaching data literacy or visual encodings, metaphoric designs enable more abstract learning and also increase humor.
% For journalists, MetaGlyph helps them to create figurative and understandable visualization easily and quickly.
}

Looking into the future, we hope our work can inspire further research from the perspectives of design expressiveness and perceptual effectiveness with regard to different visual representations (e.g., storyline~\cite{tang2020PlotThread}, data comics, and visual analytics systems~\cite{wu2022defence}).
%% Empirical study
First, we plan and also encourage researchers to conduct empirical studies to investigate how different users understand metaphoric visualizations.
Cognitive experiments can likewise be designed to derive metrics for the effectiveness of metaphor selection, so that more theories about computational metaphors can be constructed.
In turn, the theories can be utilized to inspire the development of future visualization tools.
%% Model 
Second, research on automatic visualization generation~\cite{deng2022dashbot} can benefit from large-scale visualization datasets (e.g., VisImages~\cite{Deng2020VisImages}). 
Future work can focus on constructing glyph datasets to facilitate the use of end-to-end deep learning models for an efficient generation.